\documentclass[prl, twocolumn, superscriptaddress, eprintnumbers, amsmath,amssymb]{revtex4-2}
\usepackage[]{graphicx}
\usepackage{epstopdf}
\usepackage{bm}
\usepackage{framed}

\usepackage{dcolumn}% Align table columns on decimal point
\usepackage{xcolor}

\begin{document}
\preprint{APS/123-QED}
\title{Fluctuation-Controlled Asymmetric Kinetics in Metal–Insulator Transitions}

\author{Tapas Bar} \email{tapas.bar@icn2.cat}
\affiliation{Catalan Institute of Nanoscience and Nanotechnology (ICN2), CSIC and BIST,
 Campus UAB, 08193 Bellaterra, Barcelona, Spain}
 
 \author{David Pesquera}
\affiliation{Catalan Institute of Nanoscience and Nanotechnology (ICN2), CSIC and BIST,
 Campus UAB, 08193 Bellaterra, Barcelona, Spain}
 
\author{Timm Swoboda}
\affiliation{Catalan Institute of Nanoscience and Nanotechnology (ICN2), CSIC and BIST,
 Campus UAB, 08193 Bellaterra, Barcelona, Spain}
 
\author{Arnau Villalobos-Martin}
\affiliation{Catalan Institute of Nanoscience and Nanotechnology (ICN2), CSIC and BIST,
 Campus UAB, 08193 Bellaterra, Barcelona, Spain}
\affiliation{Department of Physics, Facultat de Ciències, Universitat Autònoma de Barcelona, 08193 Bellaterra, Barcelona, Spain}
 
\author{Cristian Rodriguez-Tinoco}
\affiliation{Catalan Institute of Nanoscience and Nanotechnology (ICN2), CSIC and BIST,
 Campus UAB, 08193 Bellaterra, Barcelona, Spain}
\affiliation{Department of Physics, Facultat de Ciències, Universitat Autònoma de Barcelona, 08193 Bellaterra, Barcelona, Spain}

\author{Pol Lloveras}
\affiliation{Grup de Caracterizació de Materials, Departament de Física, EEBE and Barcelona Research Center in Multiscale Science and Engineering, Universitat Politècnica de Catalunya, Eduard Maristany, 10-14, 08019 Barcelona, Catalonia, Spain}%

\author{Marianna Sledzinska}
\affiliation{Catalan Institute of Nanoscience and Nanotechnology (ICN2), CSIC and BIST,
 Campus UAB, 08193 Bellaterra, Barcelona, Spain}

\author{Javier Rodriguez-Viejo} \email{javier.rodriguez@icn2.cat}
\affiliation{Catalan Institute of Nanoscience and Nanotechnology (ICN2), CSIC and BIST,
 Campus UAB, 08193 Bellaterra, Barcelona, Spain}
\affiliation{Department of Physics, Facultat de Ciències, Universitat Autònoma de Barcelona, 08193 Bellaterra, Barcelona, Spain}
%\date{\today}

\begin{abstract}
We report asymmetric kinetics in thermally driven metal–insulator transitions (MITs) in 1T-TaS$_2$. Using combined transport, calorimetric, and Raman measurements, we show that the transition proceeds via burst-like avalanches during cooling, while remaining continuous during heating. Although bulk transport is masked by percolative conduction, local probes and thermal measurements reveal intrinsic asymmetry in the transformation pathways. Using controlled nonequilibrium thermal perturbations generated by pulsed Joule heating, we demonstrate that the phase-ordering dynamics remains strongly athermal during cooling, whereas during heating fluctuations progressively overcome nucleation barriers, leading to a smooth transformation. The distinct responses to thermal perturbations indicate different degrees of athermality of the two hysteresis branches, which govern the transformation pathways and give rise to the observed kinetic asymmetry. These results establish a general framework in which the degree of athermality controls pathway selection in first-order phase transitions.
\end{abstract}
\maketitle

Understanding the kinetic pathways of first-order phase transitions (FOPTs) remains a central challenge in condensed matter physics \cite{Book_Slezov2009}. A large class of materials exhibiting thermally induced hysteretic transitions—including metal–insulator transitions (MITs) \cite{Chandni_prl09, Bar_prb21}, martensitic transitions \cite{Planes_prl01,Vives_prl04}, and metamagnetic transitions \cite{Samanta_prb14}—behaves athermally, where long-range interactions originate energy barriers much larger than thermal fluctuations and give rise to complex transformation kinetics. Although such transitions are governed by free-energy landscapes, the role of athermality in selecting transformation pathways—continuous versus avalanche-like—remains poorly understood. Asymmetric dynamics between forward and reverse transformations are widely observed in MITs, martensites, and metamagnetic systems \cite{Uhlir_NatCom16,Valle_prb18,Fan_prb11,Bar_prb21}. Whether asymmetric kinetics originates from unequal degrees of athermality of the two transformation branches remains largely unexplored. Previous explanations have largely attributed this asymmetry to percolative transport, based on observations in systems such as vanadium oxides and FeRh, where the resistivity evolves smoothly along one branch of the hysteresis but exhibits abrupt jumps along the other \cite{Uhlir_NatCom16,Valle_prb18,Fan_prb11}. Because transport is strongly influenced by percolation, such asymmetry is most prominently observed in nanowires or microstripes, whereas in bulk crystals the jumps are often smeared out. This has led to the prevailing view that asymmetric kinetics is primarily a consequence of low dimensionality and percolative transport \cite{Valle_prb18}.

However, asymmetric kinetics is also observed in bulk and epitaxial athermal materials using probes that are insensitive to percolation, including thermal, acoustic, optical, and spectroscopic measurements \citep{Toth_prb14, Niemann_prb12, Gallardo_prb10, Bar_prb21, Massey_prb20, Kang_ASS14, Vries_apl14, Ievgen_apl18}. In polycrystalline or disordered systems, transformation steps are often smeared by heterogeneous nucleation at grain boundaries or disorder sites \cite{Bar_prb21, Kang_ASS14}. In contrast, low-disorder systems, particularly single crystals, frequently exhibit asymmetric phase evolution during heating and cooling as an intrinsic feature of first-order phase transitions across a wide range of athermal materials \cite{Toth_prb14, Niemann_prb12}. These observations suggest that asymmetric kinetics in MITs cannot be explained solely by dimensionality, percolation, or extrinsic disorder, and instead point to an intrinsic mechanism associated with the transformation dynamics itself. Direct experimental evidence linking asymmetric kinetics to unequal degrees of athermality, however, remains lacking.

Here, we investigate the origin of asymmetric transformation pathways using combined thermal, Raman, and transport measurements on layered 1T-TaS$_2$ crystals. We find that the transition proceeds through burst-like avalanches during cooling, whereas it remains continuous during heating. Thermal measurements directly reveal jerky latent-heat propagation during cooling, while Raman spectroscopy captures the asymmetric evolution of vibrational modes across the transition. Although percolative transport masks this behavior in bulk resistance measurements, the intrinsic asymmetry becomes evident in thin layers. To probe its microscopic origin, we further examine the nonequilibrium phase-ordering dynamics under controlled thermal perturbations generated by pulsed Joule heating. We find that the heating branch responds strongly to the perturbation and evolves continuously toward the metallic phase, whereas the cooling branch exhibits only local stochastic switching and some of them relaxes back to its metastable state, indicating a much stronger degree of athermality. These results provide direct experimental evidence that unequal degrees of athermality of the two hysteresis branches govern the transformation pathways and give rise to the observed kinetic asymmetry.

The layered correlated material 1T-TaS$_2$ exhibits a sequence of temperature-driven first-order phase transitions associated with charge-density-wave (CDW) states of different commensurability. Upon cooling from the high-temperature metallic phase ($T > 550$ K), the system first enters an incommensurate CDW (I-CDW) state, followed by a nearly commensurate CDW (NC-CDW) phase at 350 K, and finally undergoes a metal–insulator transition (MIT) near 180 K into the commensurate CDW (C-CDW) phase \cite{Wang_NatCom20,Jarc_Nat23}. Upon heating, the reverse transition occurs near 220 K and proceeds through an intermediate trigonal phase up to $\sim$280 K \cite{Wang_SciRep19}. The MIT is accompanied by lattice distortions and pronounced hysteresis, making 1T-TaS$_2$ an ideal platform for investigating asymmetric transformation kinetics. In this work, we study high-quality 1T-TaS$_2$ single crystals (purity $>$ 99.995\%, HQ Graphene) during both cooling and heating cycles.

\begin{figure}
\centering
\includegraphics[width=0.47\textwidth]{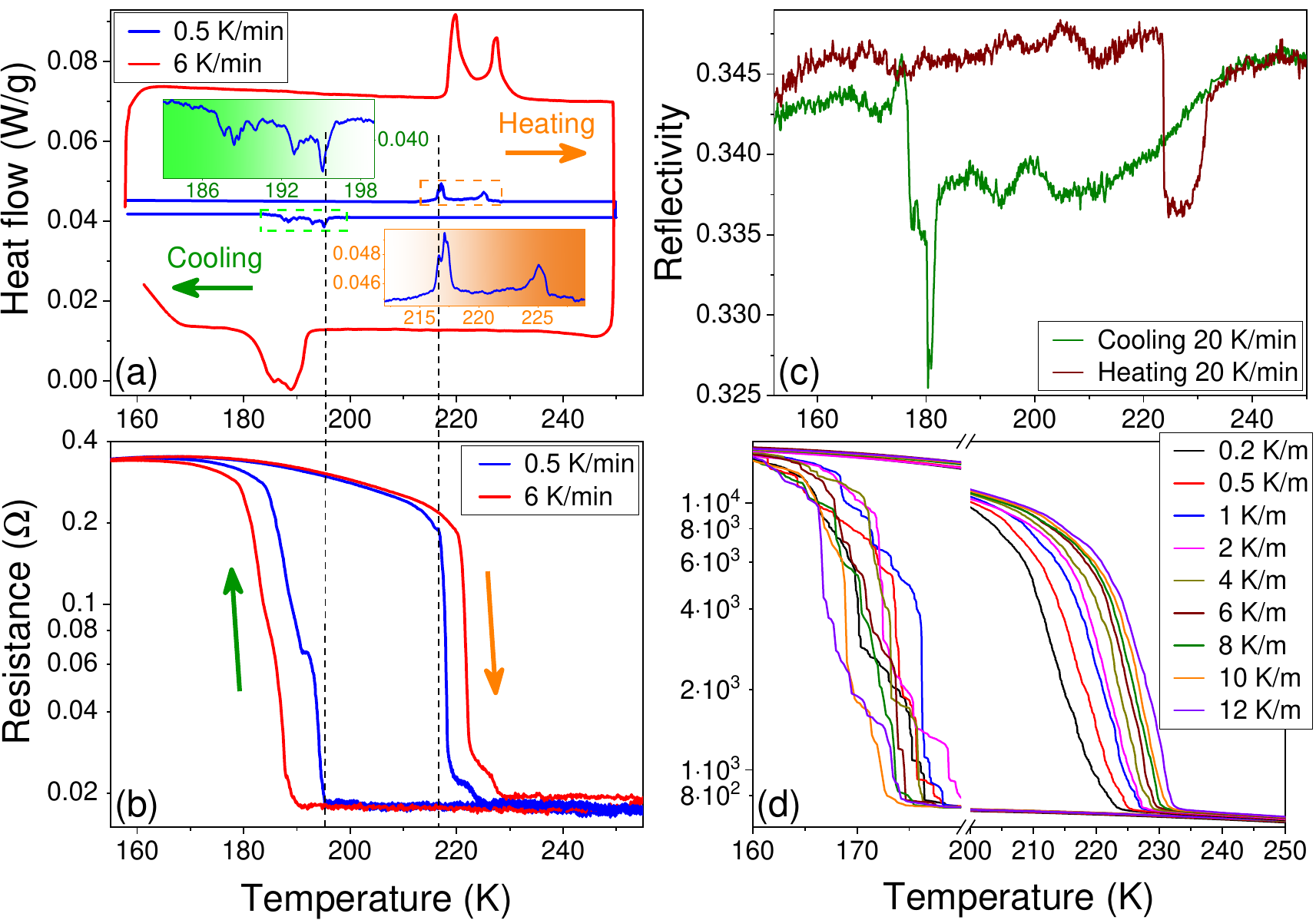}
\caption[0.5\textwidth]{(a) DSC traces of a 1T-TaS$_2$ single crystal measured during heating and cooling at scanning rates of 0.5 K/min and 6 K/min. (Inset) Expanded view of the endothermic and exothermic peaks at the slower rate. (b) Temperature-dependent resistance measured under identical ramp conditions, showing corresponding shifts in transition temperatures. (c) Reflectivity measurements revealing sharper, avalanche-like transitions during cooling compared to smoother behavior during heating. (d) Temperature-dependent resistance of a thin layer (thickness $d =$ 30 nm), demonstrating pronounced asymmetric kinetics at reduced dimensionality. The data collectively demonstrate a pronounced asymmetry between cooling and heating pathways, which becomes more evident in local probes and reduced dimensions.}
\label{fig:fig1}
\end{figure}

Figure \ref{fig:fig1}(a,b) shows differential scanning calorimetry (DSC) and transport measurements on bulk 1T-TaS$_2$ crystals acquired at two linear temperature ramps. A systematic shift of the transition temperature with ramp rate is observed, consistent with the dynamical renormalization commonly reported in hysteretic phase transitions \citep{Bar_prl18,Bar_prb23}. At slower ramp rates, both measurements reveal a two-step transformation, indicating the presence of a long-lived metastable state. A key observation is the pronounced asymmetry between cooling and heating: during cooling, the transition proceeds through multiple discrete steps accompanied by jerky latent-heat absorption, whereas during heating it evolves smoothly over a broader temperature interval [Fig. \ref{fig:fig1}(a)]. This asymmetry becomes obscured at higher ramp rates, where the experimental timescale is shorter than the characteristic relaxation time of the metastable states \cite{Gallardo_prb10}, preventing the resolution of individual transformation events, as observed at 6 K/min.

Local probes, such as reflectivity measurements using a focused laser beam, capture the avalanche-like nature of the transition during cooling much more clearly than during heating, even at temperature ramp rates nearly two orders of magnitude higher than those used in DSC measurements [Fig. \ref{fig:fig1}(c)]. The reduction in reflectivity across the transition can be attributed to enhanced light scattering arising from the coexistence of nearly commensurate (NC-CDW) and commensurate (C-CDW) domains \cite{Frenzel_prb18, Lysenko_Jap15}. Such scattering may also be viewed as a manifestation of critical opalescence in charge-density-wave transitions, where the growth of order-parameter fluctuations leads to strong diffuse scattering of light \cite{Bansal_prl24}.

In transport measurements, the asymmetric kinetics are not observed in bulk materials. However, the intrinsic behavior is revealed in thin layers \cite{Yoshida_SciAdv15, Yoshida_SciRep14} (thickness $d$ = 30 nm): during cooling, the transformation proceeds via avalanches, whereas during heating it is continuous and spans a broader temperature range [Fig. \ref{fig:fig1}(d)]. The absence of asymmetric kinetics in bulk arises from out-of-plane percolation-dominated charge transport in thick samples \cite{Martino_NPJ20, Boix_ACSNano21}. The hysteretic MIT involves the coexistence of metallic and insulating domains, with conduction governed by a small number of low-resistance percolation paths. In bulk crystals, three-dimensional percolation allows out-of-plane metallic pathways to dominate the transport response \cite{Tiwari_apl26} and mask the intrinsic transformation dynamics. In contrast, the two-dimensional resistance of thin layers is more sensitive to changes in the domain configuration, thereby revealing the asymmetric kinetics [Fig. \ref{fig:fig1}(d)]. The MIT in 1T-TaS$_2$, arising from interlayer stacking, disappears in ultrathin flakes ($d < 30$ nm), possibly because of modifications to the electronic band structure \cite{Yoshida_SciAdv15}, supercooling\cite{Chatzigiannakis_SciRep26}, or surface oxidation \cite{Tsen_PNAS15}. Notably, the transition temperature shifts systematically during heating, whereas the onset of the transition during cooling is stochastic. The observed resistance avalanches originate from the switching of one or a few metallic domains, revealing the stochastic nature of the cooling branch \cite{Sharoni_prl08}. While some degree of stochasticity is inherent in both hysteretic isothermal phase transitions \cite{Pol_NatComm19} and disordered systems \cite{Zapperi_prl97}, the asymmetric behavior points to a distinct underlying mechanism.

Although asymmetric kinetics is most readily observed in low-dimensional transport measurements \cite{Valle_prb18}, its origin is fundamentally linked to the structural transformation accompanying the MIT. Raman spectroscopy provides a direct probe of the lattice state, as the appearance, disappearance, splitting, broadening, and frequency shifts of vibrational modes are closely associated with structural and electronic changes \cite{Gasparov_prb02}. Consequently, the evolution of Raman-active phonons offers a sensitive means of tracking the phase transformation and its kinetics.

In the C-CDW phase ($T < 180$ K), 12 neighboring Ta atoms contract toward a central Ta atom to form the characteristic $\sqrt{13}a \times \sqrt{13}a \times 13c$ Star-of-David cluster \cite{Spijkerman_prb97, Gasparov_prb02}. Although the symmetry of the reconstructed unit cell remains under debate, recent studies favor the $P\bar{3}$ structure with trigonal or hexagonal stacking \cite{DjurdjiMijin_prb21}. The associated Brillouin-zone folding gives rise to a rich phonon spectrum \cite{DjurdjiMijin_prb21, Albertini_prb16, Gasparov_prb02, Hirata_SSC01}, and at 163 K we resolve at least 13 Raman-active modes in the 50–425 cm$^{-1}$ range \cite{supplementary}. Raman spectra were acquired using a 2400 mm$^{-1}$ grating and a 532 nm excitation laser focused to a spot size of approximately 500 nm. The incident laser power was limited to 0.32 mW to minimize local heating effects. In contrast, the high-temperature I-CDW phase generally exhibits only a few optical modes, two of which (E$g$ and A${1g}$) are Raman active \cite{DjurdjiMijin_prb21, Albertini_prb16}. The intermediate NC-CDW phase contains signatures of both C-CDW and I-CDW order, which are reflected in the Raman spectra \cite{supplementary, DjurdjiMijin_prb21, Albertini_prb16, Gasparov_prb02, Hirata_SSC01, He_prb16}. While the vibrational properties of the CDW phases have been extensively studied, the Mott transition itself, occurring within a narrow temperature interval ($< 5$ K), remains comparatively unexplored \cite{Gasparov_prb02}. The persistence of C-CDW signatures within the NC-CDW phase allows us to track the MITs using Raman spectroscopy, providing a direct probe of the structural evolution across the MIT.

\begin{figure}
\centering
\includegraphics[width=0.47\textwidth]{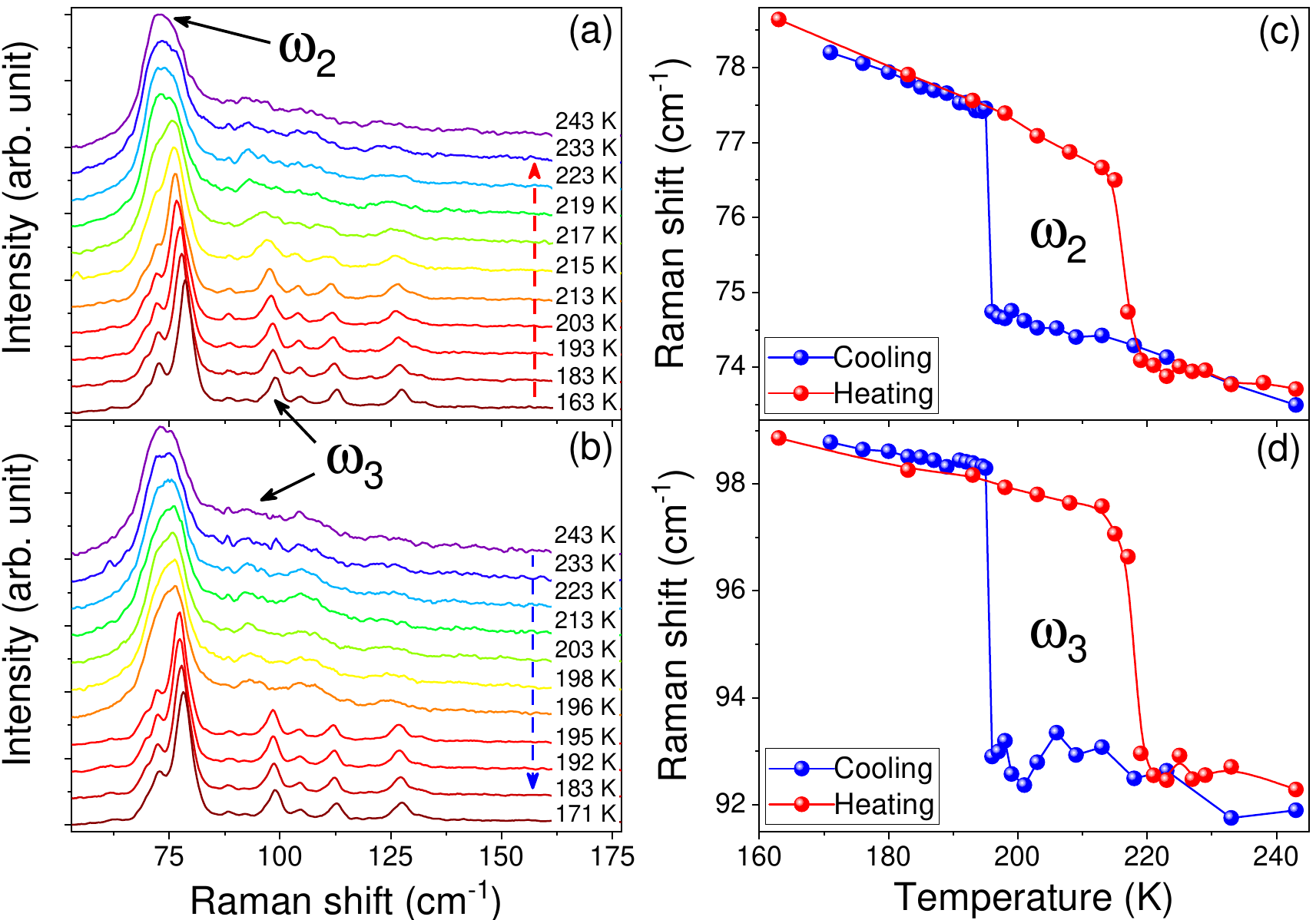}
\caption[0.5\textwidth]{Temperature-dependent low-frequency Raman spectra of bulk 1T-TaS$2$ during (a) heating and (b) cooling. (c, d) Temperature dependence of the central frequencies of the $\omega_2$ and $\omega_3$ Raman modes, showing distinct evolution during heating and cooling. Measurements were performed under quasi-isothermal conditions ($\Delta T < 0.1$ K), with a temperature sweep rate of 0.2 K/min between successive temperature set points near the transition.}
\label{fig:Temp_Raman}
\end{figure}

During the transition, the Raman modes associated with the C-CDW phase exhibit abrupt frequency shifts accompanied by a reduction in intensity [Fig. \ref{fig:Temp_Raman}(a,b)]. A pronounced kinetic asymmetry is evident in the temperature evolution of these phonon modes [Fig. \ref{fig:Temp_Raman}(c,d)]: during cooling, the transition occurs within a narrow temperature interval of approximately 1 K and is nearly single-step, whereas during heating it evolves continuously over several kelvin. The nearly single-step transformation during cooling results from the local Raman probe, which interrogates only one or a few domains. Some zone-boundary modes, such as the $\omega_1$ mode at 73 cm$^{-1}$ (see supplementary \cite{supplementary}), disappear completely across the transition and display markedly different evolution during cooling and heating \cite{Gasparov_prb02}.

The asymmetric transformation kinetics is clearly captured by the Raman modes $\omega_2$ and $\omega_3$ [Fig. \ref{fig:Temp_Raman}(c, d)], which are used throughout this work as representative probes of the structural transition. Similar behavior is observed in other low-frequency phonon modes ($<150$ cm$^{-1}$), which predominantly involve Ta vibrations \cite{DjurdjiMijin_prb21,Hirata_SSC01,He_prb16}. Higher-frequency modes, associated with vibrations of the entire S–Ta–S sandwich, also exhibit the same heating–cooling asymmetry \cite{DjurdjiMijin_prb21, Hirata_SSC01}. Interestingly, the mode at 229 cm$^{-1}$ hardens with increasing temperature, in contrast to the softening observed for most other modes \cite{supplementary}. Nevertheless, irrespective of whether a phonon mode softens or hardens, the asymmetric kinetics remains clearly discernible, demonstrating that it is an intrinsic feature of the phase transformation rather than a consequence of a particular vibrational mode \cite{supplementary}.

\begin{figure}
\centering
\includegraphics[width=0.47\textwidth]{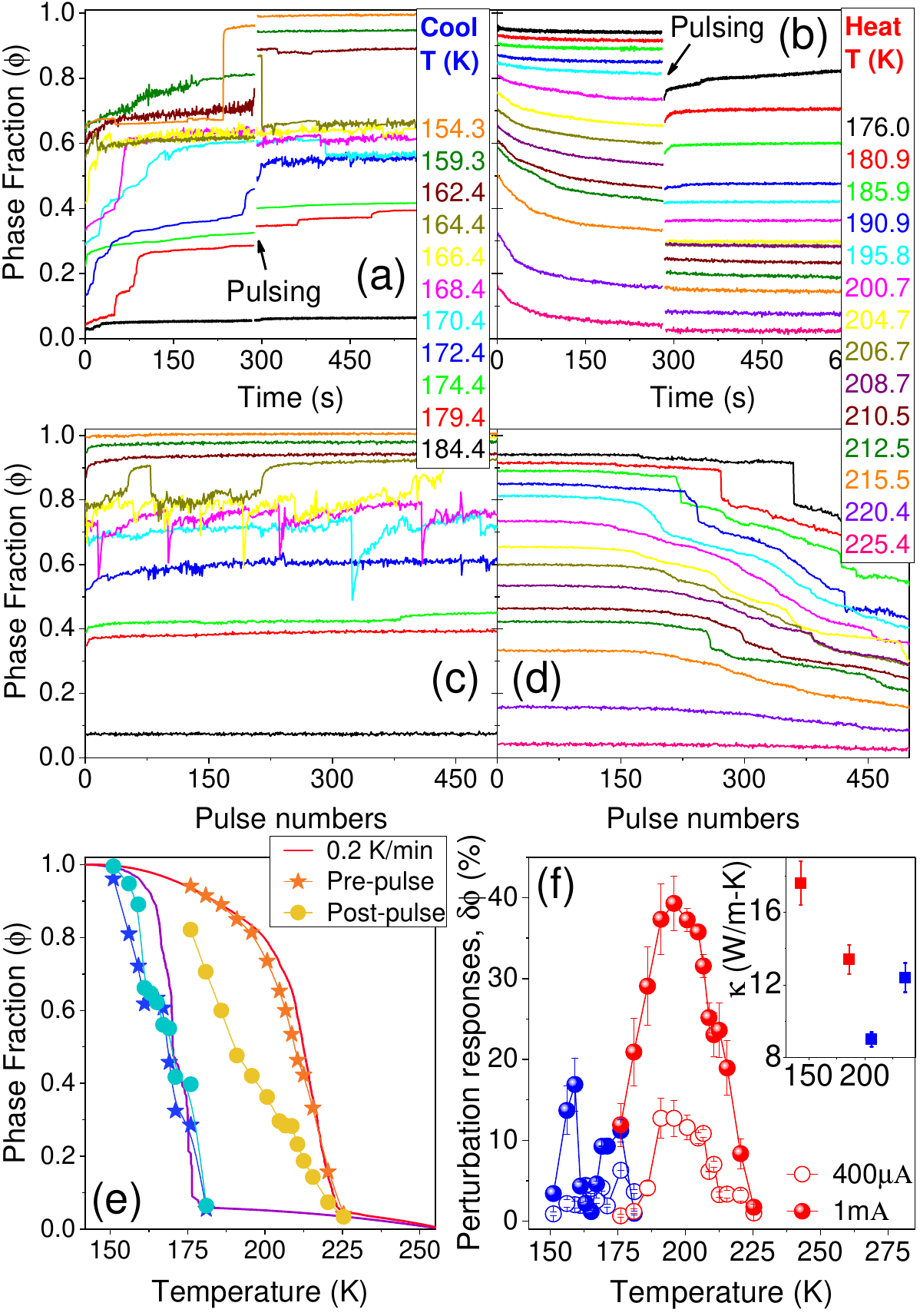}
\caption[0.5\textwidth]{Phase-ordering dynamics at different temperatures during isothermal cooling (a) and heating (b) before and after externally induced local thermal perturbations introduced at $t = 290$ s by applying 500 current pulses across a 30 nm thick layer. The temperature and pulsing profiles can be found in the End Matter. The evolution of the phase fraction as a function of pulse number during the pulsing process is shown in (c) and (d). The data reveal distinct responses to nonequilibrium thermal perturbations controlled by the pulsing currents ($I_{On}/I_{Off} = 1 mA/10\mu A$) during the cooling and heating. (e) Pre-pulsing ($\star$) and post-pulsing ($\bullet$) steady-state phase fractions, $\phi_{Pre}^{S}$ and $\phi_{Post}^{S}$, as a function of isothermal temperature. Thick lines represent the hysteresis measured under a linear temperature ramp of 0.2 K/min. (f) The total perturbation responses, $\delta \phi (\%) = (\phi_{Post}^{S}-\phi_{Pre}^{S}) \times 100$, for two different strengths of thermal perturbations. The data are shown for high ($\bullet$) and low ($\circ$) thermal perturbations corresponding to $I_{On}/I_{Off} = 1 mA/10\mu A$ and $400\mu A/10\mu A$, respectively (Heating: red; cooling: blue). (Inset) thermal conductivity of bulk 1T-TaS$_2$ during heating (red) and cooling (blue) as a function of temperature.}
\label{fig:PulseOrdering}
\end{figure}

To probe the origin of the kinetic asymmetry, we investigated the phase-ordering dynamics under controlled nonequilibrium thermal perturbations. Starting from a single-phase temperature, the sample was heated or cooled at 10 K/min to selected temperatures near the transition and subsequently held under isothermal conditions while the resistance evolution was monitored. After sufficient relaxation, the system reached a metal–insulator coexistence state, referred to as the pre-pulsing steady state, $\phi_{Pre}^{S}$ \cite{Bar_prl18, Bar_prb21}. Thermal perturbations were then introduced through Joule heating by applying millisecond current pulses across the sample, while the phase evolution was continuously tracked between successive pulses. The heating and cooling pulsing profiles can be found in the End Matter. After the pulsing sequence, the system was allowed to relax again to a post-pulsing steady state, $\phi_{Post}^{S}$ \cite{YMa_prb19}. The insulating phase fraction $\phi$, extracted using a two-dimensional percolation model \cite{Bar_prl18, Kim_prl00, Tiwari_apl26}, is shown as a function of time [Fig. \ref{fig:PulseOrdering}(a,b)] and pulse number [Fig. \ref{fig:PulseOrdering}(c,d)]. Details of the measurement protocol and phase-fraction analysis are provided in the End Matter. The pre- and post-pulsing steady-state phase fractions are compared with the phase fraction obtained under continuous temperature driving in Fig. \ref{fig:PulseOrdering}(e).

The phase-ordering dynamics exhibits a pronounced asymmetry between cooling and heating. During cooling, the phase fraction evolves through discrete avalanches [Fig. \ref{fig:PulseOrdering}(a)], whereas during heating it changes continuously [Fig. \ref{fig:PulseOrdering}(b)], providing a direct manifestation of the asymmetric transformation pathways. Under thermal perturbations, the heating branch responds strongly: the phase fraction evolves unidirectionally toward the metallic phase, indicating that the perturbation continuously assists the transformation [Fig. \ref{fig:PulseOrdering}(d)]. In contrast, during pulsing on the cooling branch, the phase evolution exhibits stochastic back-and-forth switching and does not accumulate into a complete transformation [Fig. \ref{fig:PulseOrdering}(c)]. Even near the middle of the transition, the system largely recovers from the perturbation and relaxes back to the metastable state once the pulsing is removed [Fig. \ref{fig:PulseOrdering}(a)]. The net perturbation-induced phase transformation, quantified by $\delta\phi$ [Fig. \ref{fig:PulseOrdering}(f)], is nearly twice as large during heating as during cooling.

In thermally driven first-order transitions, the system can bypass the equilibrium transition point and enter a supersaturated metastable state \cite{Book_MetaLiq20}. When long-range interactions suppress the effect of thermal fluctuations, the system remains trapped indefinitely in metastable minima, resulting in persistent hysteresis even under steady-state conditions \cite{Bar_prl18, Bar_prl20, Bar_prb21, Bar_prb23, Book_MetaLiq20, Book_Bertotti98}, as observed here [Fig. \ref{fig:PulseOrdering}(e)]. Such behavior is characteristic of athermal transitions, in contrast to conventional thermally activated transitions, such as the liquid–gas transition, where thermal fluctuations drive supersaturated states toward equilibrium according to Maxwell construction \cite{Book_Bertotti98, Book_Nonthermal15, Bar_prb21, Bar_prb23, Bar_prr25}. The distinct responses of the two hysteresis branches to thermal perturbations therefore reveal different degrees of athermality within the same transition. The cooling branch remains strongly athermal, with perturbations affecting only a limited fraction of domains before the system relaxes back to metastable configurations, whereas the heating branch is comparatively less athermal and evolves progressively toward the metallic phase under continued perturbation. A phenomenological $\phi^6$ Landau free-energy model with time-dependent mean-field dynamics, presented in the End Matter, supports this distinction between strongly athermal and fluctuation-assisted transformation kinetics. These results indicate that unequal degrees of athermality govern the transformation pathways and ultimately give rise to the observed kinetic asymmetry.

A possible origin of the unequal athermality may be related to the anomalous thermal transport that accompanies the Mott transition in strongly correlated electronic systems \cite{Lee_Science17,Liu_AdvSci20,NunezRegueiro_prl85}. As shown in Fig. \ref{fig:PulseOrdering}(f) (inset), the total thermal conductivity is higher in the insulating phase than in the metallic phase. The measured values in bulk crystals are also larger than those reported for thin layers, consistent with the previously observed thickness dependence of thermal transport in 1T-TaS$_2$ \cite{Liu_AdvSci20,NunezRegueiro_prl85,Rayappan_Physica10}. Such behavior is unusual and has been attributed to strong electronic correlations that can cause electronic heat transport to dominate over phonon-mediated transport, leading to anomalous thermal conductivity near the Mott transition \cite{Lee_Science17,Liu_AdvSci20}. Although the microscopic connection remains to be established, the pronounced difference in thermal transport between the two phases may influence how externally induced thermal perturbations propagate through the system, thereby contributing to the unequal degrees of athermality observed during heating and cooling.

Such anomalous thermal transport, sometimes associated with departures from the Wiedemann–Franz expectation, is not unique to 1T-TaS$_2$ but has also been reported in other correlated MIT systems \cite{Lee_Science17,Andreev_PSS78,Kang_jap98}. Transition-metal oxides such as VO$_2$, V$_2$O$_3$, and V$_3$O$_5$ exhibit higher thermal conductivity in the insulating phase than in the metallic phase \cite{Andreev_PSS78} and similarly display asymmetric transition kinetics \cite{YMa_peb22}.

More generally, we show that unequal degrees of athermality between the two hysteresis branches govern the selection of continuous and avalanche-like transformation pathways in first-order phase transitions. Our results provide direct experimental evidence that controlled thermal perturbations can distinguish the athermal characters of MITs. Beyond their fundamental significance, these findings offer promising prospects for neuromorphic applications \cite{Chen_AdvSci25}. In particular, the cooling branch, governed by burst-like avalanche dynamics, can emulate stochastic spiking and switching events, whereas the smoother heating branch provides a gradual analog-like response. Such intrinsic asymmetry between the two hysteresis branches may enable tunable, history-dependent switching functionalities for neuromorphic computing.

\section*{Acknowledgments}

We acknowledge Grant No. TED2021-131363B/I00 funded by MICIU/AEI/ 10.13039/501100011033 and by the European Union NextGenerationEU/PRTR. T.B. acknowledges Juan de la Cierva post-doctoral fellowship, Grant No. JDC2022-049886-I funded by MICIU/AEI/ 10.13039/501100011033 and by the European Union NextGenerationEU/PRTR. D.P. acknowledges the “Ramon y Cajal” Senior Research Fellowship (RYC-2023-044643-I) from the Spanish Ministry of Science, Innovation and Universities.  and grant PID2022-140589NB-I00, funded by MCIN/AEI/10.13039/501100011033. M.S. and T.S. acknowledge support from MICIU project ELEMENTAL (PID2023-152783OB-I00) and project PETITE  (PCI2023-143399) funded by MCIN/AEI/10.13039/501100011033 and by the European Union. The ICN2 is funded by the CERCA program/Generalitat de Catalunya. The ICN2 is supported by the Severo Ochoa program, grant  CEX2021-001214-S. T.B. J.R.-V acknowledge support from 2021SGR-00644 funded by AGAUR.

\bibliography{Sample_TaS2}

\newpage
\onecolumngrid 
\begin{center}
    \textbf{\large End Matter}
\end{center}
\twocolumngrid 

\appendix
\section{Phase-ordering measurements}

\begin{figure}
\centering
\includegraphics[width=0.47\textwidth]{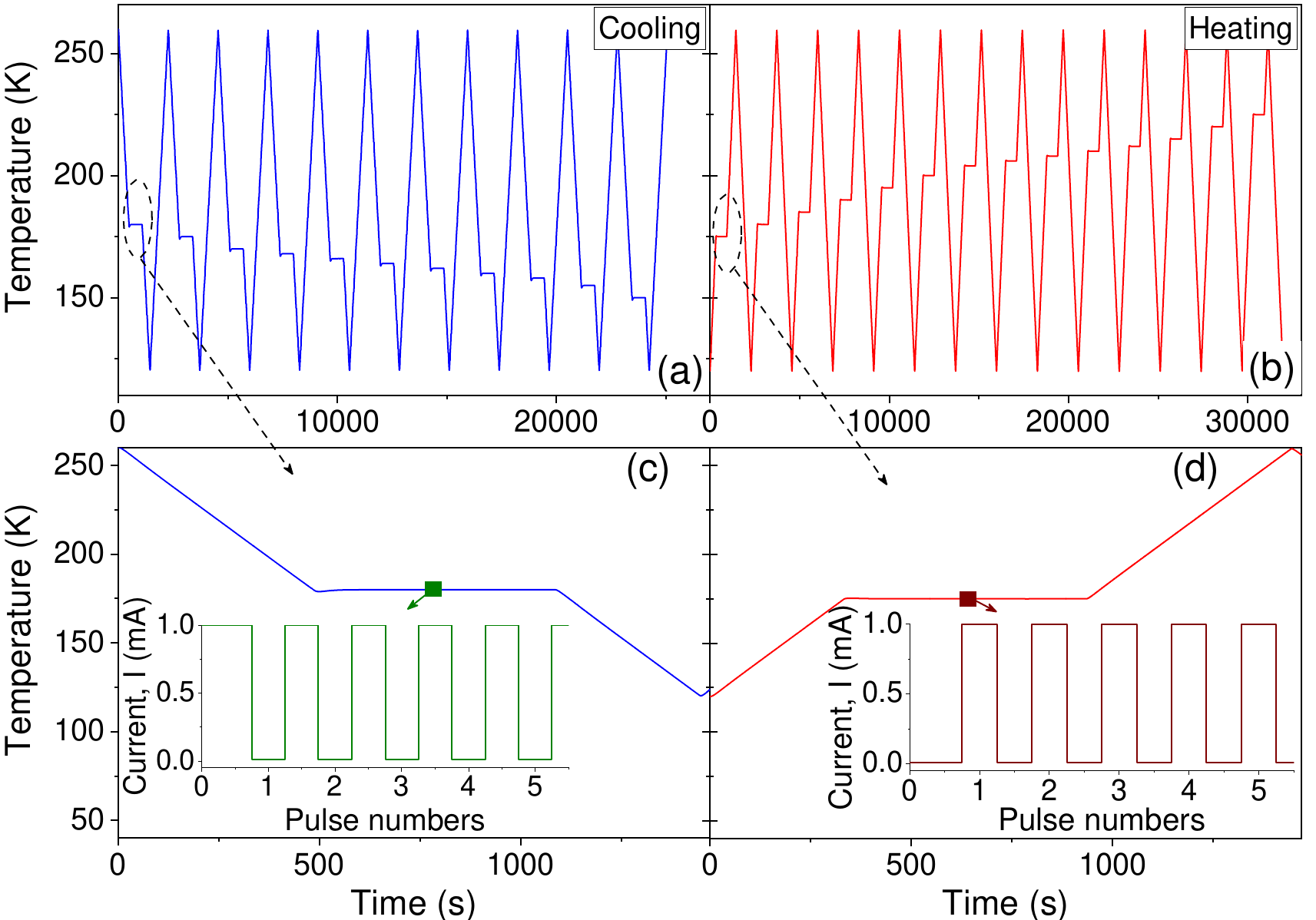}
\caption[0.5\textwidth]{Temperature profile of the phase-ordering measurements during cooling (a) and heating (b). At each isothermal temperature, the system was perturbed by applying 500 current pulses approximately 290 s after reaching the set point. Panels (c) and (d) show the corresponding pulsing profiles ($1\mathrm{mA}/10\mu\mathrm{A}$) for cooling and heating perturbations, respectively. The pulse duration and off-time were 1 ms.}
\label{fig:Temp_Profile}
\end{figure}

Each isothermal phase-ordering measurement was performed within a complete thermal cycle to erase thermal memory effects [Fig. \ref{fig:Temp_Profile}(a, b)]. Starting from temperatures far from the transition region, the sample was heated or cooled to a selected isothermal set point and allowed to equilibrate for approximately 5 min while monitoring the resistance evolution. The system was then externally perturbed by applying current pulses [Figs. \ref{fig:Temp_Profile}(c, d)], and the resistance evolution during pulsing was recorded as a function of pulse number and converted into the insulating phase fraction [Figs. \ref{fig:PulseOrdering}(c, d)]. After 500 consecutive pulses, the system was allowed to relax for an additional 5 min. The corresponding phase-ordering dynamics at different isothermal temperatures are shown in Figs. \ref{fig:PulseOrdering}(a, b).

We used a percolation model to evaluate the insulating phase fraction $\phi$ from the measured resistance data. McLachlan’s general effective medium theory \cite{McLachlan_JPC87, McLachlan_jap93} has been widely employed to estimate the phase fraction in MITs of two- and three-dimensional systems \cite{Kim_prl00} even for 1T-TaS$_2$ \cite{Tiwari_apl26}. The phase fraction $\phi$ satisfies
\begin{equation}\label{eq:percolation}
\phi \frac{(\sigma_I^{1/t} - \sigma_E^{1/t})}{(\sigma_I^{1/t} + A\sigma_E^{1/t})} + (1-\phi)\frac{(\sigma_M^{1/t} - \sigma_E^{1/t})}{(\sigma_M^{1/t} + A\sigma_E^{1/t})} = 0.
\end{equation}
where $\sigma_I$ and $\sigma_M$ are the conductivities of the insulating and metallic phases, respectively, and $\sigma_E$ is the experimentally measured conductivity. For two-dimensional systems, the critical exponent is taken as $t = 1.3$ \cite{Berg_pre24, Berg_pre23}. Unlike in three dimensions, the parameter $A = (1-\phi_c)/\phi_c$ in 2D is nonuniversal and depends on the sample geometry and correlations \cite{McLachlan_JE00}. We use $\phi_c = 0.5$ for all calculations \cite{Tiwari_apl26}; varying $\phi_c$ does not affect the qualitative behavior or the main conclusions of this work.

However, the quantitative values of $\phi$ depend on the choice of $\phi_c$. The variation of the perturbation response $\delta \phi$ for $0.3 \leq \phi_c \leq 0.6$ is shown as error bars in Fig. \ref{fig:PulseOrdering}(f). Even without employing a percolation model, the resistance change induced by external perturbations qualitatively captures the perturbation response for different pulsing amplitudes \cite{supplementary}. The results show that, for the same perturbation strength, the response during heating is approximately twice that during cooling [Fig. \ref{fig:PulseOrdering}(f)]. Phase-ordering data under weaker thermal perturbations are also provided in the Supplemental Material \cite{supplementary}.

\section{Time-dependent mean-field dynamics}
To understand the role of thermal fluctuations in the MIT, we perform simulations based on a Landau $\phi^6$ free-energy model exhibiting a temperature-dependent first-order transition. In the absence of fluctuations, the free energy at temperature $T$ is given by
\begin{equation}\label{eq_freeEng}
F(\phi, T) = \frac{a}{2}(T - T_c)\phi^2 + f_4 \phi^4 + f_6 \phi^6,
\end{equation}
where $a$, $f_4$, and $f_6$ are material parameters. A first-order transition requires $f_4 < 0$ and $f_6 > 0$ to ensure stability of the free energy \cite{Book_Chaikin95, Book_Chowdhury00}. In equilibrium, the phase fraction (order parameter $\phi$) is obtained by minimizing Eq.~(\ref{eq_freeEng}). Under external driving (temperature or field), the order parameter evolves according to the time-dependent Ginzburg–Landau (TDGL) equation \cite{Book_Chaikin95, Bar_prl18, Bar_prl20, Zhong_prl05}:
\begin{equation}\label{Eq_TDGL}
\partial_t \phi = -\tau \frac{\delta F(\phi, T)}{\delta \phi},
\end{equation}
where $\tau$ is the characteristic relaxation time.

\begin{figure}
\centering
\includegraphics[width=0.47\textwidth]{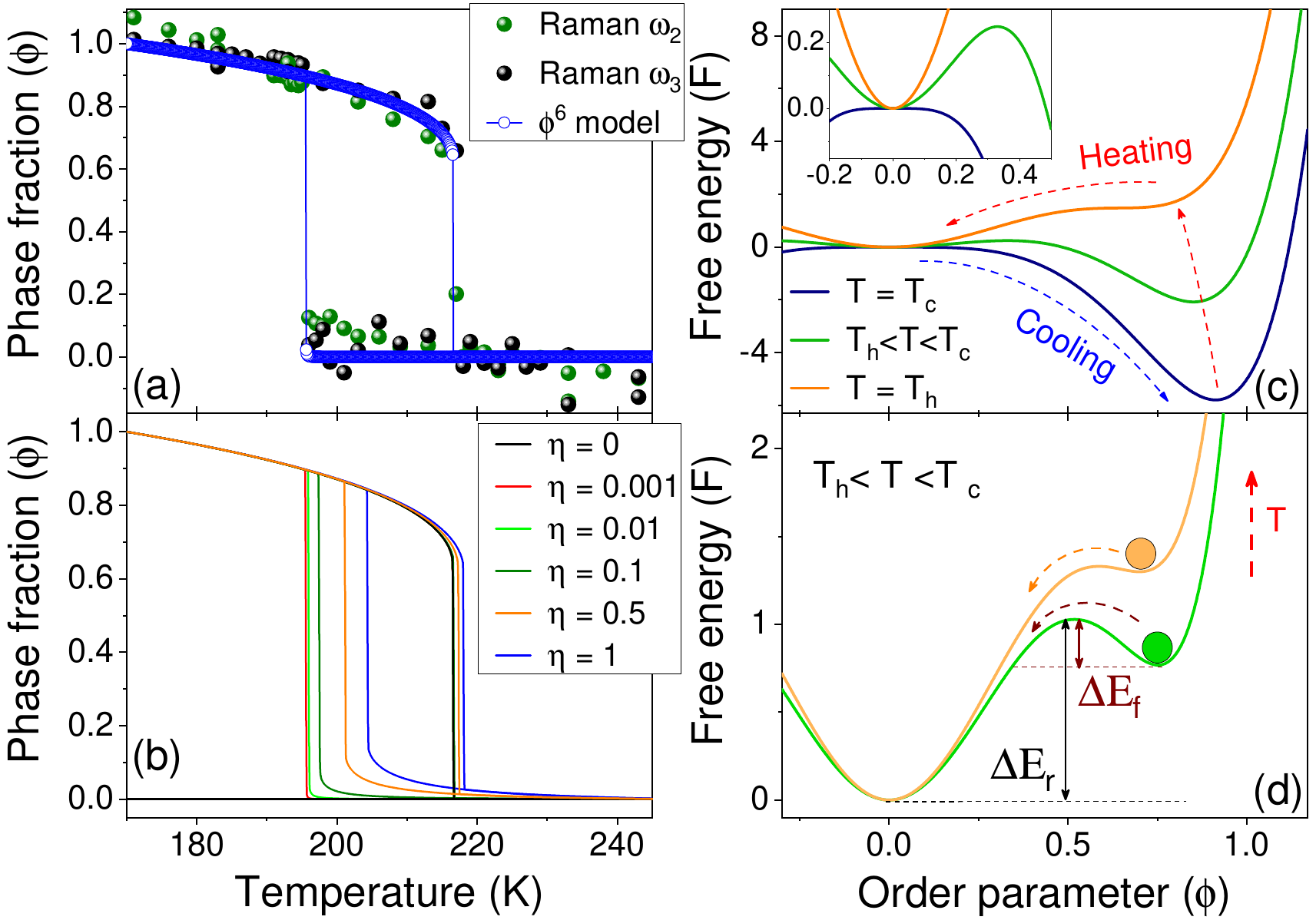}
\caption[0.5\textwidth]{Normalized Raman shifts of the $\omega_2$ and $\omega_3$ modes, representing the insulating phase fraction $\phi$ (a). The phase fraction obtained from the Landau $\phi^6$ free energy [Eq. \ref{eq_freeEng}] is overlaid with the experimental data. (b) Landau $\phi^6$ free energy $F(\phi)$ at representative temperatures, illustrating the evolution of metastable states; $T_c$ and $T_h$ denote the steady-state transition temperatures during cooling and heating, respectively. (c) Temperature dependence of the order parameter $\phi$ for different noise strengths, calculated using TDGL dynamics [Eq. \ref{Eq_TDGL_Noise}]. (d) Schematic of barrier-crossing phase-ordering dynamics when the perturbation strength becomes comparable to the free-energy barrier ($\Delta E_f$) before the spinodal point. Stronger fluctuations induce earlier transformations, leading to a reduction of the hysteresis width.}
\label{fig:Theory}
\end{figure}

Under steady-state conditions, the metastable nucleation barrier of the $\phi^6$ free energy vanishes at two temperatures, $T_c$ and $T_h$. In the absence of fluctuations (athermal), the system transforms at these limiting temperatures, corresponding to spinodal instabilities \cite{Book_Chaikin95, Bar_prl18, Bar_prl20}. Under quasi-static temperature driving, the free energy evolves via TDGL dynamics and follows the steady-state path \cite{Bar_prl18, Zhong_pre95}. The transformation during cooling occurs at $T_c$, while the transition temperature during heating, $T_h$, is governed by the parameters $a$, $f_4$, and $f_6$. For bulk 1T-TaS$_2$, $T_c = 195.5$ K, and we choose $a = 0.5$, $f_4 = -25$, and $f_6 = 20$ to reproduce experimental value $T_h = 216.5$ K. Figure \ref{fig:Theory}(a) shows that the temperature dependence of the order parameter obtained from the Landau free energy follows the normalized Raman modes.

At finite temperature, real materials are subject to fluctuations arising from thermal, field, or disorder effects. These fluctuations influence forward and reverse transformations differently due to the asymmetric free-energy landscape \cite{Book_Berglund06}. They are incorporated as a Langevin noise term in the TDGL equation \cite{Ma_prl26, Ma_pre26, Kundu_pre23}:
\begin{equation}\label{Eq_TDGL_Noise}
\partial_t \phi = -\tau \frac{\delta F(\phi, T)}{\delta \phi} + \eta (r, t),
\end{equation}
where $\eta(r,t)$ is a stochastic field distributed in space and time.

To capture the essential features of asymmetric kinetics, we perform TDGL simulations using an effective noise amplitude defined as $\eta = \langle \eta (r, t) \rangle$. Figure \ref{fig:Theory}(b) shows that increasing noise strength rounds the sharp transition.

At $T = T_h$, the double-well free-energy potential ($\phi = \phi_M = 0$ and $\phi = \phi_I$) collapses into a single-well potential, rendering the metastable phase unstable and the transformation inevitable [Fig. \ref{fig:Theory}(c)]. In contrast, during cooling at $T \leq T_c$, an unstable fixed point exists at $\phi = 0$, and even a small fluctuation is sufficient to drive the system away from this point [Fig. \ref{fig:Theory}(b)]. In the absence of fluctuations ($\eta = 0$), the system remains trapped at this unstable fixed point and does not transform from the metallic to the insulating phase [Fig. \ref{fig:Theory}(b), black line]. The shift in hysteresis toward higher temperature has been understood as a dynamical alteration of fluctuations into an effective field. In thermally driven transitions, fluctuations act as an effective field, introducing a renormalized term $\delta H$ [$\delta H \sim \eta(r,t)/\tau$] in the free-energy expression [Eq. \ref{eq_freeEng}], thereby shifting the free-energy landscape toward higher temperatures.

The key point is that the nature of the transition is governed primarily by the strength of the perturbation rather than its microscopic origin. As the perturbation strength increases, a sharp avalanche-like transition progressively evolves into a smoother and more continuous transformation, accompanied by a reduction of the hysteresis width \cite{Ma_pre26, Kundu_pre23}. Consequently, the more strongly athermal branch of the hysteresis exhibits sharper kinetics than the fluctuation-affected branch.

In long-range interacting systems, most local fluctuations are suppressed, and external perturbations can trigger nucleation only when their magnitude becomes comparable to the metastable free-energy barrier ($\Delta E_f$) \cite{Book_MetaLiq20, Book_Bertotti98, YMa_apl20}. For weak perturbations, only a small fraction of the sample undergoes local switching, while stronger perturbations promote multiple nucleation and growth events, resulting in a progressive transformation. Increasing perturbation strength also enables barrier crossing further away from the spinodal point, thereby reducing the hysteresis width, consistent with the behavior observed during heating [Fig. \ref{fig:PulseOrdering}(e)].

The numerical results suggest that the heating branch of the MIT in 1T-TaS$_2$ is less athermal owing to the stronger influence of thermal fluctuations, whereas the cooling branch remains predominantly athermal. Thus, the Landau $\phi^6$ free-energy model combined with TDGL dynamics captures the essential features of the fluctuation-driven asymmetric kinetics.

\newpage
\onecolumngrid 
\begin{center}
    \textbf{\large Supplemental Material}
\end{center}
\onecolumngrid

This Supplemental Material provides additional experimental details and supporting analyses for the results presented in the main text. It includes the temperature evolution of hardening and softening Raman modes, the methodology used to extract thermal conductivity from Raman measurements, and a detailed analysis of phase-ordering dynamics under thermal perturbations.

\section{Raman modes}

\begin{figure}
\centering
\includegraphics[width=0.53\textwidth]{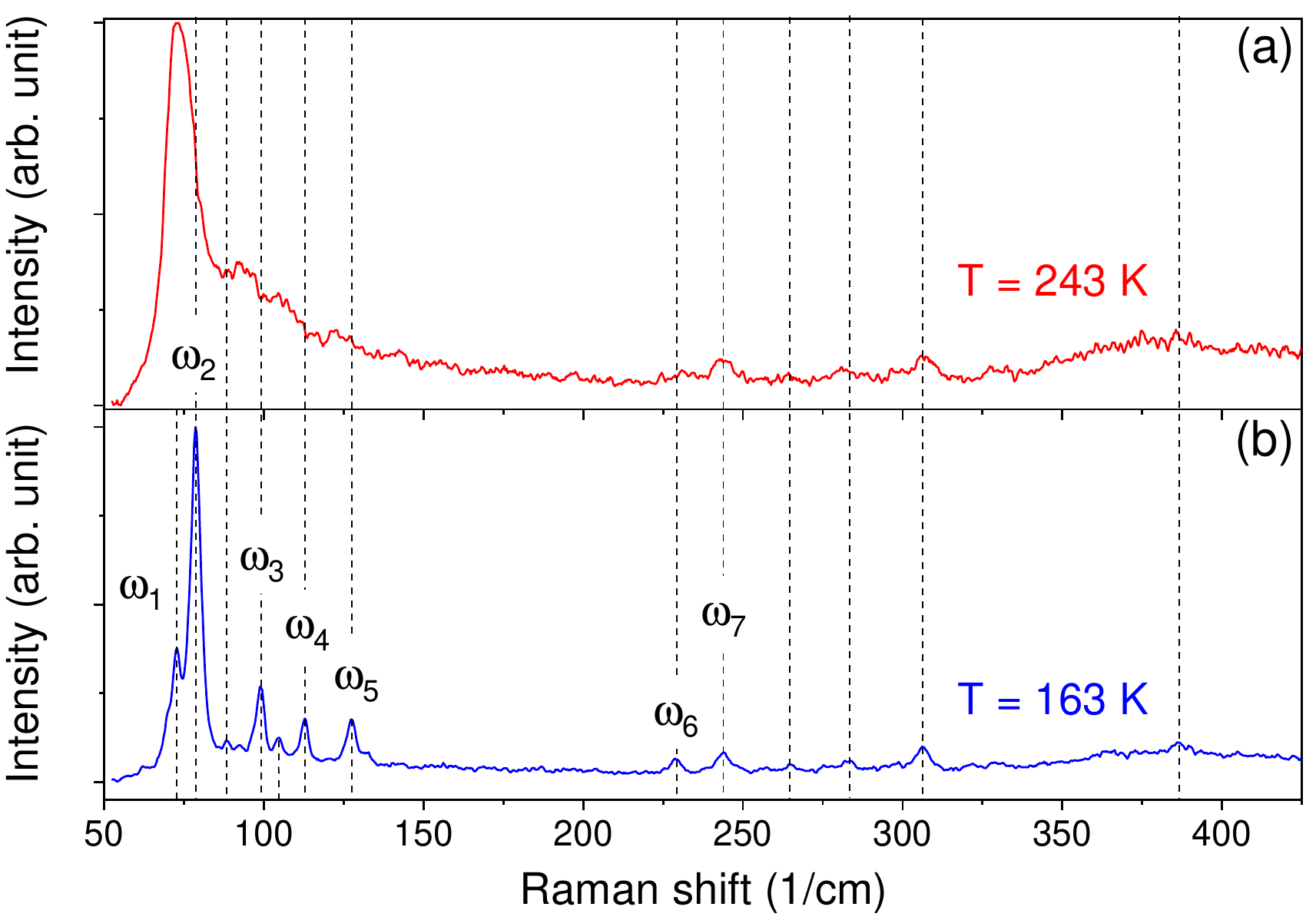}
\caption[0.5\textwidth]{Raman spectra of bulk 1T-TaS$_2$ measured at selected temperatures during heating. The Raman modes associated with the low-temperature commensurate charge-density-wave (C-CDW) phase broaden, weaken, and shift with increasing temperature, indicating the persistence of C-CDW domains within the nearly commensurate CDW (NC-CDW) phase above the insulator-to-metal transition ($T_h \approx 216.5$ K).}
\label{fig:Raman}
\end{figure}

At low temperatures, 1T-TaS$_2$ possess commensurate charge density wave (C-CDW) phase, where Ta atoms organize into characteristic Star-of-David clusters \cite{Spijkerman_prb97, Gasparov_prb02}. Factor group analysis predicts 57 (19A$_g$ + 19E$_g$, with E modes doubly degenerate) Raman-active modes and 60 (20A$_u$ + 20E$_u$) infrared-active modes \cite{Albertini_prb16, Gasparov_prb02, DjurdjiMijin_prb21, Hirata_SSC01, He_prb16}. As temperature increases, several modes broaden and overlap, reducing their experimental visibility. At $T = 163$ K, we resolve at least 13 phonon modes, some of which are labeled as $\omega_1$–$\omega_7$ in Fig. \ref{fig:Raman}(b). With increasing temperature, 1T-TaS$_2$ transforms from the C-CDW phase to the nearly commensurate CDW (NC-CDW) phase. At higher temperatures (350 K $<$ T $<$ 554 K), the incommensurate CDW (I-CDW) phase exhibits only four optical modes, of which two (E$g$ and A${1g}$) are Raman-active \cite{DjurdjiMijin_prb21, Albertini_prb16}. The NC-CDW phase consists of coexisting C-CDW and I-CDW domains, and consequently both low- and high-temperature Raman modes are observed simultaneously [Fig. \ref{fig:Raman}(a)].
 
Although the C-CDW to NC-CDW transition extends over a relatively broad temperature range, the associated metal–insulator transition is comparatively sharp. The abrupt nature of the transition is most clearly reflected in the zone-boundary modes, such as $\omega_1$, which disappears completely across the transition [see Fig. 2 of the main text].  In particular, above the metal–insulator transition, several phonon modes shift toward either lower or higher frequencies, corresponding to mode softening and hardening, respectively [Fig. \ref{fig:Raman}]. A subset of these modes was used to investigate the kinetic asymmetry during heating and cooling.

\begin{figure}
\centering
\includegraphics[width=0.53\textwidth]{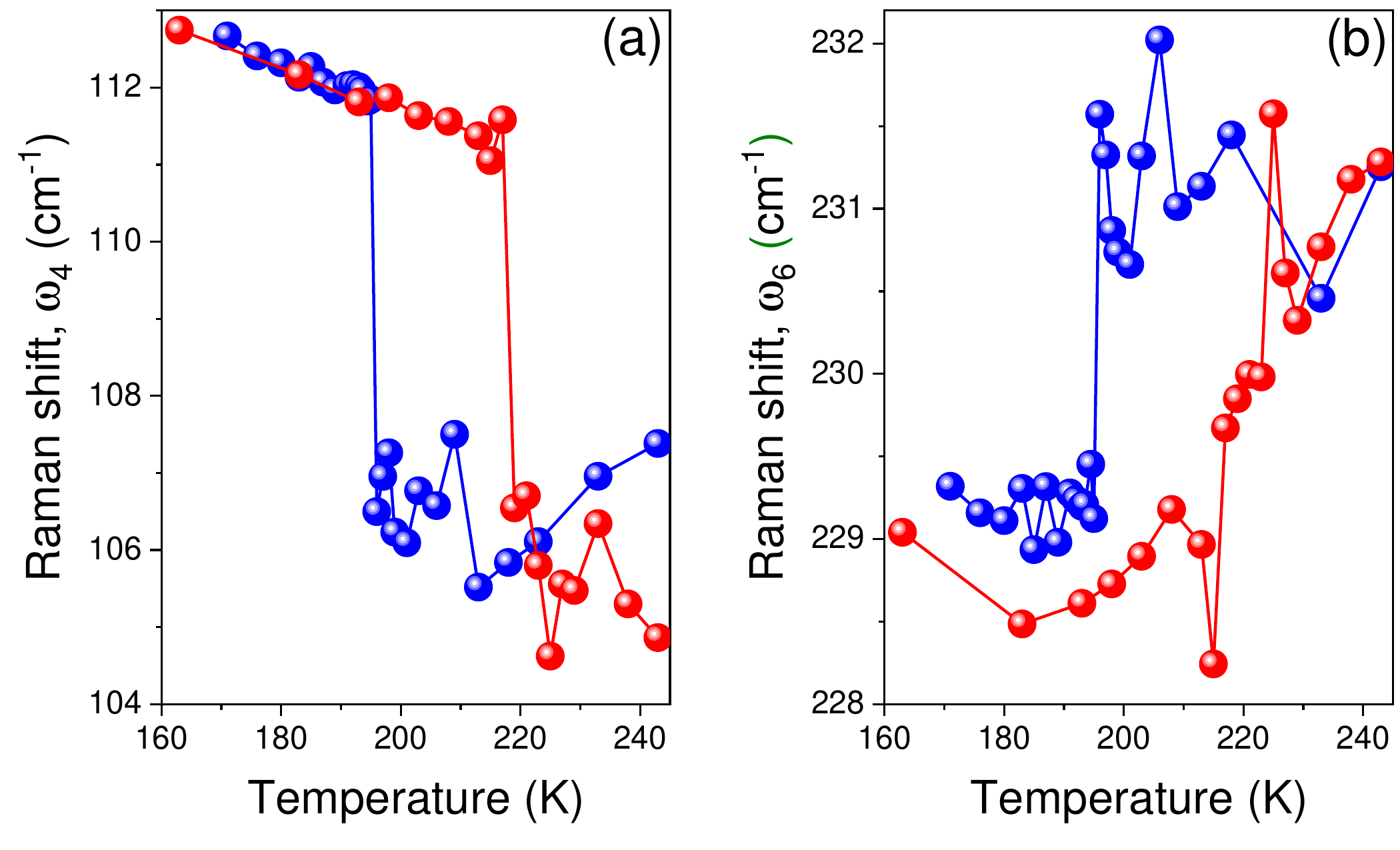}
\caption[0.5\textwidth]{(a) Temperature dependence of a softening phonon mode and (b) a hardening phonon mode during heating and cooling. Kinetic asymmetry is evident in both cases, irrespective of whether the phonon mode softens or hardens across the transition.}
\label{fig:Hardening}
\end{figure}

The temperature-dependent Raman spectra of 1T-TaS$_2$ and the corresponding frequency shifts of two representative Raman modes ($\omega_2$ and $\omega_3$) are presented in Fig. 2 of the main text. In addition to these modes, several other phonon modes also exhibit kinetic asymmetry during the forward and reverse transitions. While most modes soften with increasing temperature, a few high-frequency modes display anomalous hardening. Figure \ref{fig:Hardening} compares representative examples of mode softening and mode hardening.

Owing to the large mass difference between Ta and S atoms, the low-frequency phonon modes predominantly involve Ta vibrations, whereas the high-frequency optical modes have a significant contribution from S vibrations \cite{DjurdjiMijin_prb21, Hirata_SSC01}. The observed mode hardening may arise from changes in the effective force constants associated with the structural transition or from modifications of the electronic structure near the Fermi surface \cite{Gasparov_prb02}. Importantly, both hardening and softening modes exhibit a sharper transition during cooling than during heating. Thus, irrespective of the sign of the frequency shift, the kinetic asymmetry remains robust, indicating that it is intrinsically linked to the structural transformation accompanying the metal–insulator transition [Fig. \ref{fig:Hardening}].

\section{Thermal Conductivity}

\begin{figure}
\centering
\includegraphics[width=0.77\textwidth]{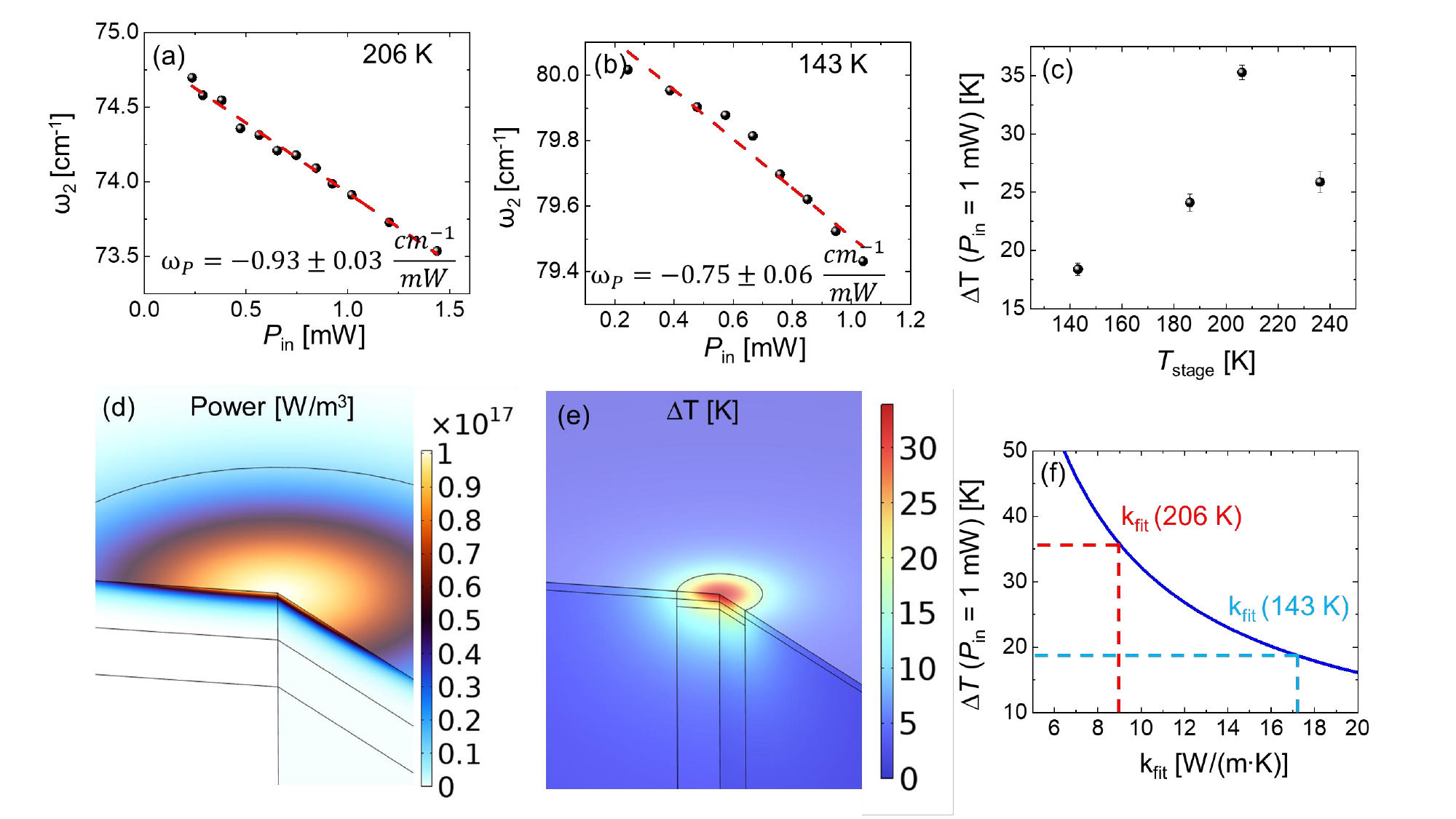}
\caption[0.5\textwidth]{(a,b) Fitted peak $\omega_2$ as a function of the incident laser power Pin at a stage temperature of (a) 206 K and (b) 143 K. (c) Average temperature increase $\Delta$T of the laser spot radius as a function of the sample stage temperature for an incident laser power of 1 mW. (d) Power map showing the power density at the heated sample. (e) Temperature map of the heated area. (f) Calculated temperature increase $\Delta$T of the laser spot radius as a function of the fitting parameter $\kappa_{fit}$}
\label{fig:figSimulation}
\end{figure}

For the characterization of the thermal conductivity, we use the Raman setup within the same configuration as in the previous measurements. In this case, we hold the sample stage at four different temperatures and locally heat the sample by increasing the laser power of the Raman laser. Figure \ref{fig:figSimulation} (a,b) shows the power series of the fitted peak $\omega_2$ obtained at 206 K and 143 K.

Similarly to the temperature measurements the heating through the increased laser power induces a red shift in the Raman peak position $\omega_2$. From figure 2(c) [main text], we extract the linear temperature dependence of $\omega_2$ , obtaining slopes of -0.04 cm$^{-1}$/K at lower temperature below the transition (during heating) and -0.025 cm$^{-1}$/K at higher temperatures, above the transition (during cooling).

The temperature increase per power is then calculated by using the power dependent shift $\omega_P$ and the temperature dependent red shift $\omega_T$ as follows,

\begin{equation} \label{Equation_X}
\Delta T = \frac{\omega_P}{\omega_T }
\end{equation}

Figure  \ref{fig:figSimulation} (c) shows the temperature increase for 1 mW of incident power at the four measured temperatures.

To quantify the results, we employ a heat transfer in solids finite element method (FEM) model implemented in the software COMSOL Multiphysics. The material is modelled as a uniform TaS$_2$ material block. The heat source is described as a depth-dependent Gaussian laser beam with the power P defined as,

\begin{equation} \label{Equation_Y}
P = \frac{2A_c}{\pi r_0^2}P_0(1-Ref)\times exp[-\frac{2r^2A_c(t-z)}{r_0^2}]
\end{equation}

where  P$_0$ is the total incident laser power, $Ref$ is the reflectance, A$_c$ is the absorption coefficient, r$_0$ is the $1/e^2$ laser spot radius, $r$ is the radial coordinate with the center of the laser spot located at $r$ = 0, $t$ is the thickness of the block, and $z$ is the vertical coordinate.  Figure \ref{fig:figSimulation} (d) shows the power source distribution in the comsol model. The power is integrated over the entire geometry giving $P_0(1-Ref)$, which verifies the validity of the model.

The absorption coefficient is set to $0.6\times10^6$ cm$^{-1}$[Table \ref{table}] based on literature values. The laser spot size of the Raman setup is measured using the knife edge method and is found to be 500 nm. The reflectance is measured directly within our Raman setup [see Fig. 1 (c) main text]. All simulation parameters are summarized in Table \ref{table}. 

The power source is applied within the first 100 nm of the sample following formula \ref{Equation_X}. At this depth, more than 99 $\%$ of the incident power is absorbed and the temperature values are corrected accordingly. We then calculate the temperature increase relative to the ambient temperature $\Delta T$, which is approximated within the laser spot radius and over the same reference thickness of 100 nm. Figure \ref{fig:figSimulation} (e) shows the calculated temperature map of  the heated sample.

Next, we sweep the thermal conductivity of TaS$_2$ $\kappa_{fit}$, assuming an isotropic behavior. The total incident laser power is set to 1 mW to match the experimentally measured temperatures values. By fitting the experimental data with the simulation results, we extract $\kappa_{fit}$ as illustrated in Figure \ref{fig:figSimulation} (f). The final fitted results are shown in main text [Fig. 3(f)(inset)]

\setlength{\tabcolsep}{10pt}

\begin{table}[h]
\centering
\begin{tabular}{lc}
\hline
\hline
Parameter & Value \\
\hline
\\[-0.7em]
Input power: $Q_0$   &   $1$ mw \\

Absorption coefficient: $A_{c}$   &   $0.6\times 10^6$ $cm_{-1}$ \\

Laser spot radius: $r_{0}$   &   $0.5$ $\mu m$\\

Thickness of block: $t$   &   $0.2$ $mm$ \\

Length of system: $L$   &   $100$ $\mu m$ \\

Thermal conductivity: $\kappa_{fit}$   &  $10$ $w/m-K$\\

Power (at r = 0, t = 0): $P_{0}$   &   $Q_0(1-Ref)(\frac{2A_c}{\pi r_0^2})$ \\

Reflectance: $Ref$   &   $0.34$ \\

Reference thickness: $t_{abs}$   &   $100$ $nm$ \\

Ambient temperature: $T_{amb}$   &   $140$ $K$ \\
\hline
\hline
\end{tabular}
\caption{Parameter of COMSOL simulations.}
\label{table}
\end{table}

\section{Phase ordering measurements}

The extent of phase transformation under external perturbations depends strongly on both the perturbation strength and its duration. When the perturbation strength is small compared to the free-energy barrier height, the system remains effectively athermal and can enter a supersaturated metastable state. In this regime, the transformation occurs only when the barrier height becomes comparable to the thermal fluctuations or external perturbations. Consequently, the steady-state phase fraction still follows a hysteretic path, although with a reduced hysteresis width compared to the unperturbed case.

In contrast, when the perturbation strength exceeds the free-energy barrier height, the system can readily overcome the metastable barrier and relax toward the global free-energy minimum. As a result, the metastable state becomes inaccessible, the transition becomes reversible, and the hysteresis progressively vanishes. Therefore, the competition between the perturbation strength and the free-energy barrier height determines the transformation kinetics. A more detailed discussion based on the Landau free-energy landscape is provided in the End Matter section of the main text.

\begin{figure}
\centering
\includegraphics[width=0.7\textwidth]{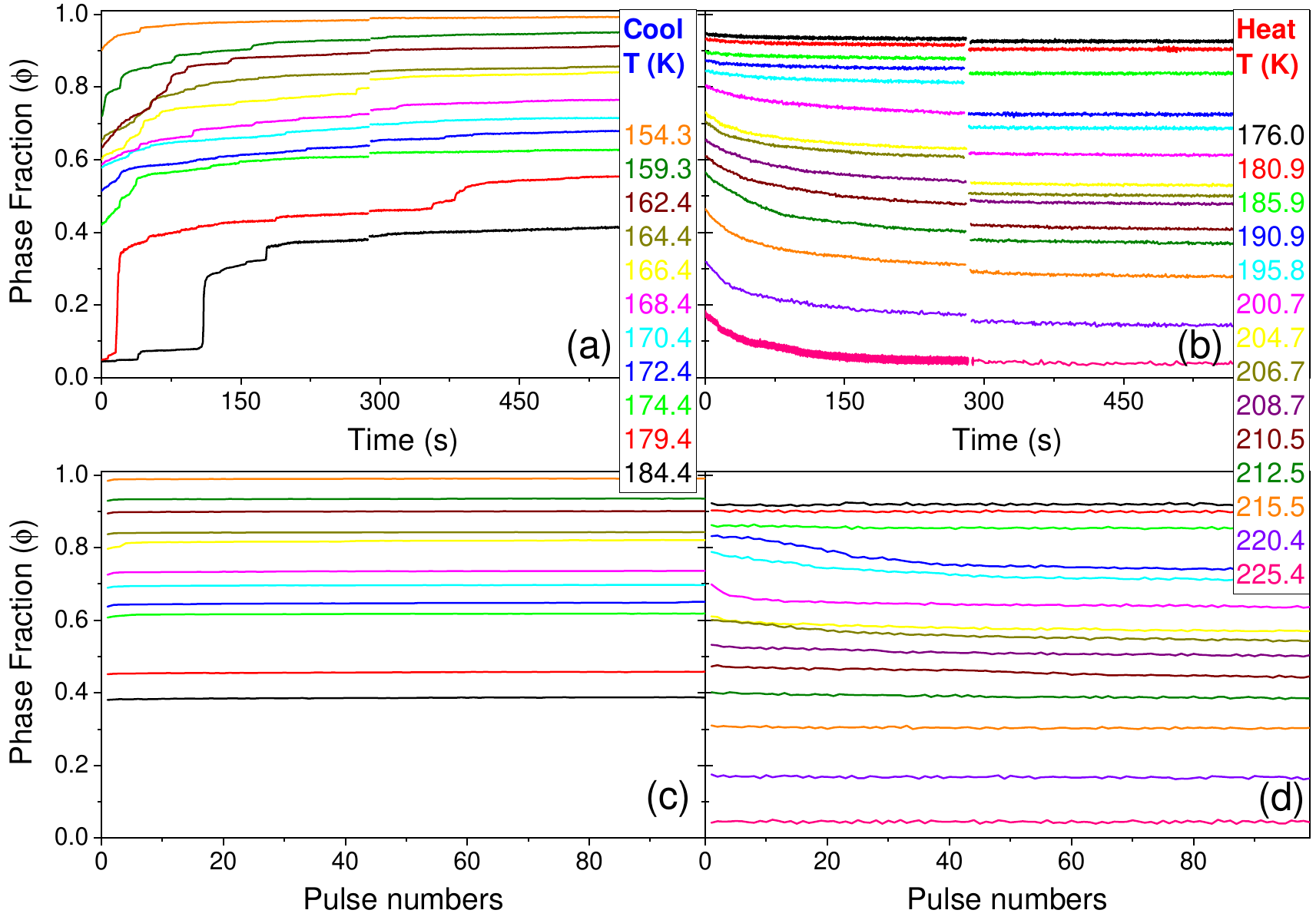}
\caption[0.5\textwidth]{Phase-ordering dynamics under lower thermal perturbations ($I_{on}/I_{off} = 400,\mu\mathrm{A}/10,\mu\mathrm{A}$). The time evolution of the phase fraction during isothermal cooling (a) and heating (b) before and after the thermal perturbation. The temperature and pulsing profiles are identical to those used for the higher perturbation measurements presented in the main text, except for the reduced pulsing current $I_{on}$. The evolution of the phase fraction as a function of pulse number is shown in (c) and (d). In contrast to the higher perturbation case, the system reaches its steady-state value within only a few pulses. The percentage difference between the pre-pulsing and post-pulsing steady-state phase fractions as a function of isothermal temperature is presented in the main text [Fig. 3(f)]. For comparison, the same quantity was also evaluated directly from the resistance data, without employing the percolation model, and is shown in Fig. \ref{fig:NoiseResponse_Res}(b).}
\label{fig:Only_400uA}
\end{figure}

Consistent with this picture, the change in phase fraction under weaker thermal perturbations [Fig. \ref{fig:Only_400uA}] is smaller than that observed under stronger perturbations [Fig. 3 of the main text]. This behavior indicates that increasing the perturbation strength progressively reduces the hysteresis width.

\begin{figure}
\centering
\includegraphics[width=0.47\textwidth]{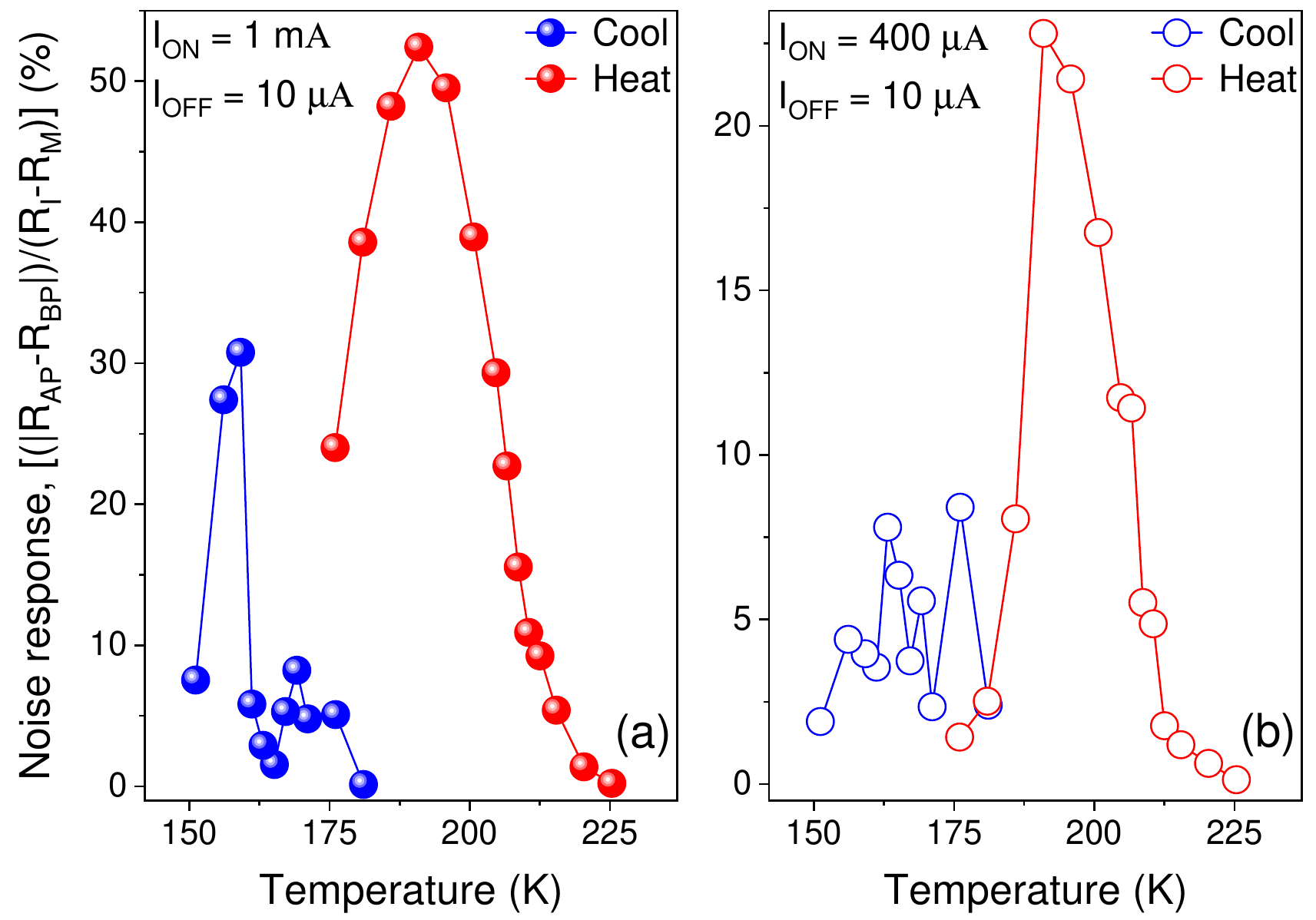}
\caption[0.5\textwidth]{Total percentage change in phase fraction calculated from the steady-state resistance values measured before ($R_{BP}$) and after ($R_{AP}$) the thermal perturbation. Here, $R_I$ and $R_M$ represent the resistances of the insulating and metallic phases, respectively. To directly compare the responses during heating and cooling, the pulse amplitude, pulse duration, and total number of pulses were kept identical for all measurements. The resulting perturbation response is shown for higher (a) and lower (b) thermal perturbation strengths.}
\label{fig:NoiseResponse_Res}
\end{figure}

The most important observation is that the response to thermal perturbations is consistently larger during heating than during cooling, irrespective of the perturbation strength [Fig. 3(f) of the main text and Fig. \ref{fig:NoiseResponse_Res}]. This asymmetry is robust and does not depend on the particular percolation model used to extract the phase fraction, as evidenced by the weak dependence on the percolation threshold shown in Fig. 3(f) of the main text. Furthermore, the perturbation response extracted directly from the resistance data exhibits the same heating–cooling asymmetry as the phase-fraction analysis, demonstrating that the central conclusion of this work is independent of the percolation model. These results demonstrate that the observed asymmetry is an intrinsic feature of the phase-ordering dynamics rather than an artifact of the phase-fraction analysis.

The asymmetric fluctuation response indicates that the insulator-to-metal transition during heating is highly sensitive to thermal perturbations, whereas the reverse transition during cooling remains predominantly athermal. Consequently, thermal perturbations can effectively drive the transformation during heating but have only a limited influence during cooling, where the dynamics are controlled primarily by the driving temperature. This unequal sensitivity to thermal perturbations underlies the asymmetric kinetics and constitutes the central finding of this work.

\end{document}